\documentstyle[emulateapj,psfig]{article}
\lefthead{Agol}
\righthead{Sgr A* Polarization}
\begin{document}

\title{Sgr A* Polarization: No ADAF, Low Accretion Rate, and Non-Thermal 
Synchrotron Emission}

\author{Eric Agol}
\affil{Physics and Astronomy Department, Johns Hopkins University,
    Baltimore, MD 21218; agol@pha.jhu.edu}
\begin{abstract}
The recent detection of polarized radiation from Sgr A* requires a 
non-thermal electron distribution for the
emitting plasma. 
The Faraday rotation measure must be small, placing
strong limits on the density and magnetic field strength.  We show that 
these constraints rule out advection-dominated accretion flow models.
We construct a simple two-component model which can reproduce both the 
radio to mm spectrum and the polarization.  This model predicts that the 
polarization should rise to nearly 100\% at shorter wavelengths.  The 
first component, possibly a black-hole powered jet, is compact, low 
density, and self-absorbed near 1 mm with ordered magnetic field,
relativistic Alfv\'en speed, and a non-thermal electron distribution. 
The second component is poorly constrained, but may be a 
convection-dominated accretion flow with $\dot M \sim 10^{-9}M_\odot/$yr,
in which feedback from accretion onto the black hole suppresses the accretion 
rate at large radii.  The black hole shadow should be detectable with
sub-mm VLBI.
\end{abstract}


\section{Introduction}

The nearest supermassive black hole candidate lies at the center of the
Milky Way galaxy, weighing in at $2.6\times10^6 M_\odot$, as inferred from
motions of stars near the galactic center (Ghez et al. 1998; Genzel et
al. 1997).  The low
luminosity of the point source associated with the center, $\la 10^{37}$ erg/s,
is a conundrum since accretion from stellar winds of neighboring
stars should create a luminosity of $\sim 10^{41}$erg/s.  One possibility
is that most of the energy is carried by the 
accreting matter into the black hole, as in the advection-dominated
accretion flow solution (ADAF, Narayan \& Yi 1994; Narayan, Yi, \& Mahadevan 1995).  
Such a situation is achieved when most of the dissipated energy is channeled into
protons which cannot radiate efficiently.   Low efficiency also occurs
when gas accretes spherically and carries its energy in as kinetic
energy (Melia 1992).  Alternatively, the accretion rate may be overestimated,
and the emission may be due to a tenuous disk or jet (Falcke, Mannheim,
\& Biermann 1993).

The radio spectrum of Sgr A* can be described by a power law, $F_\nu \propto \nu^{1/3}$
from centimeter to millimeter wavelengths.  This is intriguingly close
to the spectrum of optically thin, mono-energetic electrons emitting
synchrotron radiation (Beckert \& Duschl 1997).  However, this explanation
is not unique:  a self-absorbed source which varies in size as a
function of frequency may produce a similar spectral slope (Melia 1992;
Narayan et al. 1995).  A possible technique to distinguish
these models is to measure the polarization of the emission:  Faraday
rotation and self-absorption can change the polarization magnitude
and wavelength dependence (Jones \& O'Dell 1977).

Only recently has linear polarization been detected at high frequency by
Aitken et al. (2000: A00); previous searches at lower frequency showed only 
upper limits (Bower et al. 1999a,b).  After correcting for contamination by 
dust and free-free emission, the inferred polarization is 10-20\%, implying 
a synchrotron origin.  This correction is made somewhat uncertain by
the large beam size.  Remarkably, the polarization shows a change in position 
angle of $\sim90^\circ$ around 1 mm, which A00 suggest might be
due to synchrotron self-absorption.  We first discuss the physics of 
synchrotron polarization (\S 2); we then apply it to various models in 
the literature (\S  3); next we discuss a model consistent with all of 
the observations (\S 4); and finally speculate on the physical implications 
of this model (\S 5).

\section{Synchrotron Theory Background}

In the synchrotron limit ($\gamma \gg 1$) for an isotropic electron velocity 
distribution, some analytic results have been 
derived, which we now summarize (Ginzburg \& Syrovatskii 1965 \& 1969: GS ).  
For a uniform slab of electrons 
with a power-law distribution, $d n_e/ d \gamma \propto \gamma^{-\xi}$
(with $\gamma_{min} \le \gamma \ll \gamma_{max}$ such that electrons with
$\gamma_{min}$ and $\gamma_{max}$ do not contribute to the frequency of interest), 
we can relate the magnetic field strength and electron density 
in the slab to the fluid-frame brightness temperature and
the spectral turnover due to self-absorption.
For $\xi=2$ and a uniform field $B_\bot$ (projected into the sky 
plane) we find $B_\bot \sim 2  T_{11}^{-2} \nu_{12}$  G and
$\tau_C \sim 3\times 10^{-2} T_{11}^4 \nu_{12}\gamma_{min}^{-1}$, 
where $T_{11}$ is the brightness temperature in units of $10^{11}$ K 
at the self-absorption frequency $\nu_t = 10^{12}  \nu_{12}$ Hz and 
$\tau_C$ is the Compton scattering optical depth of the emission region.
For $\nu < \nu_t$, the emission is self-absorbed so
$F_\nu \propto \nu^{5/2}$, while above this frequency the emission
is optically-thin and $F_\nu \propto \nu^{(1-\xi)/2}\exp(-\nu/\nu_{max})$
where $\nu_{max} = 3 B_\bot e \gamma_{max}^2/(4 \pi m_e c)$.

In the optically-thin regime, the polarization plane is perpendicular to 
the magnetic field with polarization $\Pi = (\xi +1)/(\xi + 7/3)$, up to 100\% 
for $\xi \gg 1$. In the optically-thick regime,  $\Pi = -3 / (6 \xi + 13)$ (for
$\xi > 1/3$);  the radiation polarized perpendicular to the magnetic field 
is absorbed more strongly than the opposite polarization,
causing the radiation polarized along the magnetic field to dominate, 
switching the polarization angle by $90^\circ$, which changes the sign of $\Pi$.
Numerical calculations show that the optically-thick polarization peaks at 
$|\Pi| = 20$\% for $\xi=1/3$, but remains large for $0 < \xi < 2$.  

To compute the polarization near the self-absorption frequency
requires a knowledge of the polarized
opacity and emissivity, $\mu_{\bot,\|},\epsilon_{\bot,\|}$.  For 
$\xi=2$, these can be approximated as (GS):
$\mu_{\bot,\|} = r_s^{-1} (\nu/\nu_t)^{-3} (1\pm 3/4)$ and
$(\epsilon_{\bot,\|}/ \mu_{\bot,\|}) = 2S_t/9 (\nu/\nu_t)^{5/2}(13\pm9)/(4\pm1)$
where  $r_s$ is the size of the emission region, $\nu_t$ is the frequency for 
which the total source has an optical depth of unity (i.e. $\tau = \mu r_s = 
1/2(\mu_\bot+\mu_\|)r_s = 1$), $S_t$ is the source function near the 
frequency $\nu_t$, and the $+$ or $-$ signs go with the radiation emitted $\bot$ or 
$\|$ to the magnetic field, respectively.  GS then express the polarization and 
emission for a slab with uniform magnetic field strength and direction, constant 
density, and size $r_s$: $I_\bot = (\epsilon_\bot/\mu_\bot) (1-\exp(-\mu_\bot r_s))$,
$I_\|   = (\epsilon_\|/\mu_\|) (1-\exp(-\mu_\|   r_s))$, and
$\Pi = (I_\bot - I_\|)/(I_\bot + I_\|),$
where $I_\bot, I_\|$ are the intensities (erg/cm$^2$/s/Hz/sr) with polarization 
perpendicular and parallel to the projected direction of the magnetic field on 
the sky. 

For electron distributions which are highly peaked at a single
energy (such as mono-energetic or relativistic Maxwellian)  the
polarization for $\nu \la \nu_t$ is zero.

The Faraday effect rotates the polarization vector of photons emerging
from different optical depths by different amounts, causing a cancellation in 
polarization (Agol \& Blaes 1996).  The differential Faraday rotation angle within 
the source scales as $\Delta \theta = 3.6\times 10^{28} \tau_{phot} B \nu^{-2}
\gamma_{min}^{-2}$ (Jones \& O'Dell 1977), where $\tau_{phot}$ is the Compton optical
depth of the photosphere.  When optically thin,  $\tau_{phot}\sim \tau_C$ is 
constant, so rotation is largest at the self-absorbed wavelength.  When 
self-absorbed, $\tau_{phot}$ of the photosphere scales as $\nu^{\xi/2+2}$, so 
the differential Faraday rotation angle $\propto \nu^{\xi/2}$ (for  $\xi > 1/3$), 
again largest at the self-absorption wavelength.  The differential rotation
at $\nu_t$ is $\Delta \theta \sim 2\pi g(\xi) (\theta_b/\gamma_{min})^\xi/\gamma_{min}$,
where $\gamma_{min}$ is the minimum electron Doppler factor, $g(\xi)$ is a dimensionless
factor of order unity, and $\theta_b$ is
the brightness temperature in units of $m_e c^2/k_B$.

\section{Observational Constraints on Published Models}

The observations of polarization in Sgr A* provide the following
constraints on emission models:

1)  The differential Faraday rotation angle near the self-absorbed wavelength must
be $\ll \pi$.

2)  The electron distribution must be non-thermal since
the polarization due to a thermal electron distribution is 
suppressed when self-absorbed by a factor of $\exp(-\tau)$.
If the beam correction by A00 is correct, then  $\Pi \sim$ 
12\% at self-absorbed wavelengths, requiring $\xi \la 2$.

3)  The self-absorption frequency must lie near the change in
polarization angle, $\sim 1$mm.

4)  The component contributing at lower frequencies must have
zero linear polarization.

5)  The magnetic field must be ordered to prevent cancellation
of polarization.

These constraints rule out several models proposed in the
literature, as will be discussed in turn.

The low efficiency of an ADAF implies a higher accretion rate and 
thus higher density than for a high efficiency flow of the same
luminosity and geometrical thickness.  For Sgr A*, an 
accretion rate of $\sim 10^{-(4-5)} M_\odot/$yr is inferred
due to capture of gas in the vicinity of the black hole
(Quataert, Narayan, \& Reid 1999; Coker \& Melia 1999), which is the
value assumed in ADAF models.  Assuming that the gas falls
in at near the free-fall speed, one infers 
an electron density $n_e = 10^{10}$ cm$^{-3} \dot m_{-5}
x^{-3/2}$ and a magnetic field strength of
$B= 10^3 {\rm G} \dot m_{-5}^{1/2} x^{-5/4} (v_A/0.1v_{ff})$,
where $x$ is the radius of the emission region in units of
$r_g=GM/c^2$, $\dot m_{-5}$ is the accretion rate in units
of $10^{-5} M_\odot$/yr, and $v_A/v_{ff}$ is the ratio of the
Alfv\'en speed to the free-fall speed.  These values imply a total
Faraday rotation angle at the self-absorption frequency $\nu_t$ of
$\Delta \theta \sim 10^{4} \dot m_{-5}^{3/2} \nu_{12}^{-2}
(v_A/0.1v_{ff})$.
This value is so large that rotation of the emitted
radiation leads to zero net polarization, so ADAFs are in direct
conflict with the observed polarization.  Only significant
modifications of the model, such as a reduction in the accretion
rate by a factor of $10^{-3}$, can reduce the Faraday rotation
angle $\ll \pi$.  
An accretion rate of $10^{-8} M_\odot/$yr is consistent with
the observed luminosity if the accretion flow has a higher efficiency
$\sim$2\%, no longer ``advection-dominated.''
In addition, ADAF models assume a Maxwellian electron distribution,
which cannot produce the observed switch in polarization angle\footnote{
Mahadevan (1999) and \"Ozel, Psaltis, \& Narayan (2000) have added a non-thermal
electron component to ADAF models which contributes to the flux at
wavelengths longer than 2 mm, not at the polarized wavelengths.}.
Finally, ADAFs predict a higher self-absorption frequency:
\"Ozel et al. (2000) find that $\nu_t \sim 5\times10^{12} 
\dot m_{-5}^{5/9}{\rm Hz}$, which implies $\dot M \sim 4\times10^{-7} M_\odot$/yr
to be consistent with the observed $\nu_t \sim 5\times 10^{11}$Hz.
The accretion rate might be reduced if there is significant gas lost
by a wind or jet (Begelman \& Blandford 1999; Quataert \& Narayan
1999) or if the Bondi rate is reduced by heating the infalling gas
with heat carried outwards by a convection-dominated accretion flow, 
or ``CDAF'' (Igumenshchev \& Abramowicz 1999, 2000; Stone, Pringle, 
\& Begelman 1999; Quataert \& Gruzinov 2000; Narayan, Igumenshchev,
\& Abramowicz 2000; Igumenshchev, Abramowicz, \& Narayan 2000).

The model of Melia (1992) is rather similar to the ADAF model, and thus 
suffers the same problems: the high accretion rate implies high density 
which is inconsistent with the observed polarization.

Beckert \& Duschl (1997) considered several 1-zone, quasi-monoenergetic
and thermal emission models for the synchrotron emission.  These electron 
distributions do not produce a swing in polarization angle by 90 degrees
since the polarization is suppressed when self-absorbed.  Their model
does produce a self-absorption frequency near the correct frequency,
however.  Falcke, Mannheim, \& Biermann  (1993) present a disk-plus-jet 
model which assumes a tangled magnetic field topology which would erase 
any polarization.  However, an ordered magnetic field would be a small 
change to their model which might bring it into line with the polarization
observations.

\section{A Phenomenological Model}

Now, we attempt to construct a model consistent with all of the
data, using uniform emission regions for simplicity.  
Typical optically-thin AGN spectra show $\xi \sim 2-3$; since
$\xi=2$ is consistent with the polarization from A00,  we fix
$\xi=2$ in our model fits.  The model parameters
for the polarized component are 
$S_t=6$ Jy, $\nu_t = 550$ GHz (corresponding to $\lambda = 
0.55$ mm), and $\nu_{max} \sim 5000$GHz (Figure 1).

To explain the lack of polarization and spectral slope flatter than 5/2,
we require an additional component which is unpolarized 
and has a cutoff near 1 mm so that it doesn't dilute the polarization at 
shorter wavelengths.  Since Sgr A* has a spectral slope of 1/3 at mm 
wavelengths and appears to have a spectral turnover at 1 GHz, we model 
the spectrum as a monoenergetic electron distribution with energy
$\gamma$ and zero polarization (due to Faraday depolarization or
tangled magnetic field) which becomes self-absorbed at low frequency
(Beckert \& Duschl 1997).  For the unpolarized component, we find 
$F_\nu=1.3(\nu/\nu_{max})^{1/3} \exp(-\nu/\nu_{max}){\rm Jy}$ with 
$\nu_{max} \sim 50$GHz, and $\nu_t \sim 1$GHz (Figure 1).  

\vskip 2mm
\hbox{~}
\centerline{\psfig{file=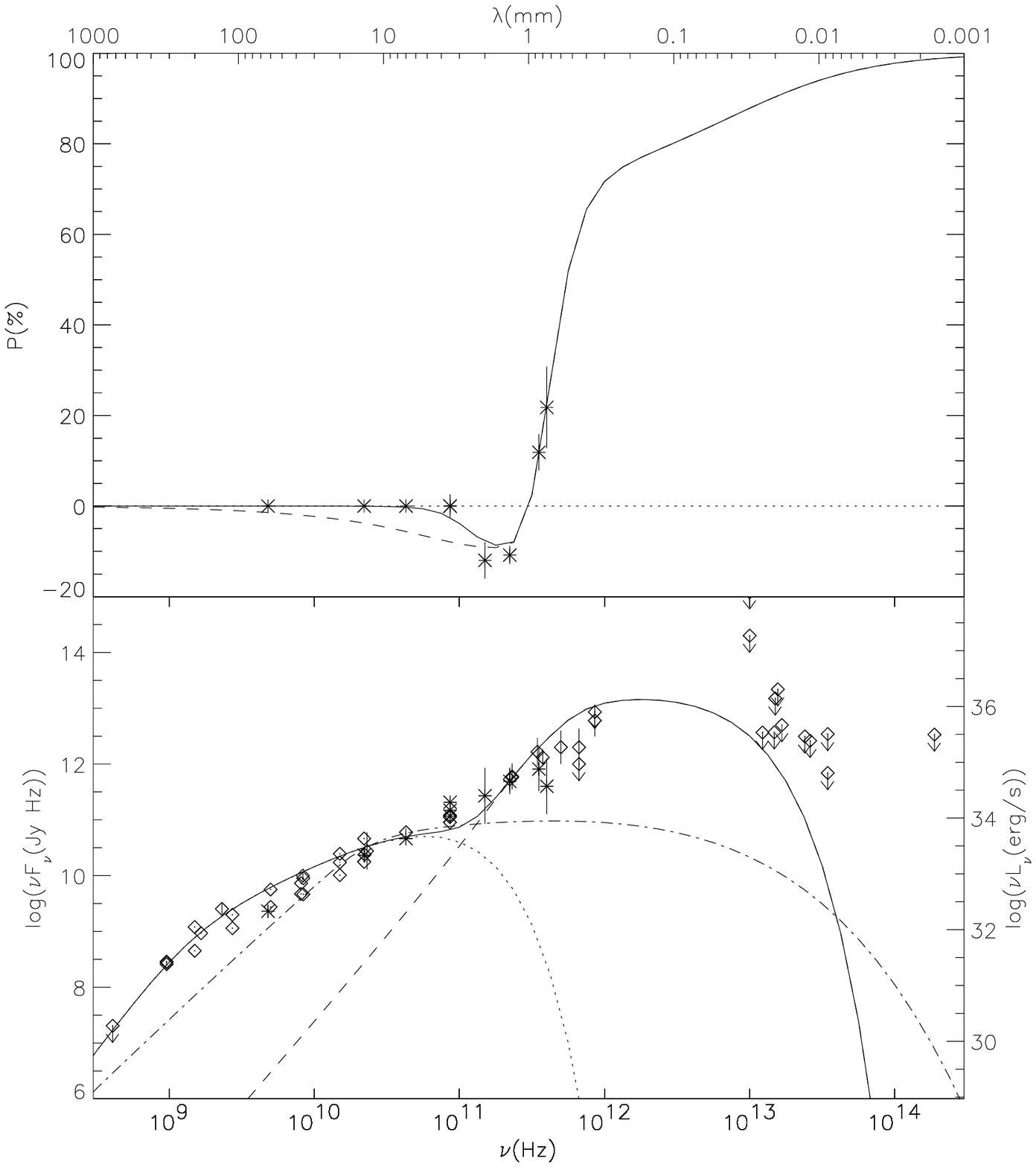,width=3.6in}} 
\noindent{
\scriptsize \addtolength{\baselineskip}{-3pt}
\vskip 1mm
\begin{normalsize}
Fig.~1: Polarization and spectral energy distribution of Sgr A*
compared to model.  The dashed line shows the polarized component,
the dotted line the unpolarized, mono-energetic component, and the
solid line the sum of the two.  The dot-dash line shows the maximum
CDAF model (assumed to be unpolarized;  the total polarization is similar
if the CDAF replaces the monoenergetic component).  The diamonds are
the data compiled by Narayan et al. (1995), while the asterisks are
the data from Bower et al. (1999a,b) and A00.
\end{normalsize}
\vskip 3mm
\addtolength{\baselineskip}{3pt}
}

Figure 1 compares the model to the data.  To compare the polarization, 
we have plotted the Stokes' parameter that lies at $83^\circ$.  Remarkably, 
the polarization should rise to $\sim 100$\% at even shorter wavelengths.

\subsection{Physical conditions in Synchrotron emitting regions}

Krichbaum et al. (1998) report a source radius of $55 \mu$as at 1.4 
mm from VLBI observations; this corresponds to 19$r_g$.
The source size may be smaller at higher frequencies,  but we expect 
the radius of the emission region to be greater than the size of the 
event horizon of the black hole, which has an apparent size of $\sim 5r_g \sim 
15 \mu$as projected on the sky (including gravitational bending,
Bardeen 1973), so we use an intermediate size in further estimates.
The flux of the fitted model at the self-absorption frequency,
$\nu_t=550$ GHz, is $\sim 9$ Jy.  This implies a brightness 
temperature in the emission frame $T_b \sim 1.6\times10^{10} (r_s/10r_g)^{-2}
\Gamma^{-1}$ K, where $r_s$ is the size of the source (we have assumed the 
area of the source is $\pi r_s^2$) and $\Gamma$ is the bulk Doppler boost 
parameter.  For a 
steeply falling electron number distribution, $k T_b \sim 4 \gamma m_e c^2$ 
(for $\xi=2$), where $\gamma m_e c^2$ is the energy of the emitting 
electrons, implying $\gamma \sim 10 (r_s/10r_g)^{-2}\Gamma^{-1}$ for the 
electrons at the self-absorption frequency.  Using the formulae from
\S 2,  we find: $B_\bot = 350  (r_s/10r_g)^4 \Gamma {\rm G}$, 
$\tau_C = 10^{-5}  (r_s/10r_g)^{-8} \Gamma^{-5}$, and 
$\gamma_{max} = 50 (r_s/10r_g)^{-2} \Gamma^{-1}$, implying
$n_e \sim 6\times 10^6 (r_s/10r_g)^{-9} \Gamma^{-5}$cm$^{-3}$.
The ratio of magnetic to rest-mass energy density is
$B^2/(8\pi n_e m_p c^2) \sim 1 (r_s/10r_g)^{17}
\Gamma^{9/2} $ for an electron-proton plasma, indicating a
relativistic Alfv\'en speed. 
The Faraday rotation angle at $\nu_t
=5.5\times10^{11}$ Hz is $\Delta \theta \sim 350
(r_s/10r_g)^{-4} \Gamma^{-2} \gamma_{min}^{-3}$, assuming $B_\| \sim B_\bot$.
For $r_s \sim 10 r_g$, $\gamma_{min}$ can
be as large as 4, reducing $\Delta \theta$ to 5; for $r_s \sim 5 r_g$,
$\gamma_{min}$ can be as large as 20 reducing $\Delta \theta$ to $\sim 0.6$.
Alternatively, if the synchrotron emission is due to a pair plasma, 
Faraday rotation will be reduced by the ratio of the proton number
density to the pair number density.
The rotation angle is further reduced at the observed wavelengths by
a factor $\sim\nu/\nu_t$.  The high energy cutoff for the electron
distribution may be due to synchrotron cooling since 
$t_{cool} = 8\times10^8 \gamma_{max}^{-1} B^{-2}
\sim 6 (r_s/10r_g)^{-6}\Gamma^{-3}$ sec, similar to
the dynamical time, $t_D \sim 13 x^{-3/2}$ sec.
Given the strong scaling of quantities with the unknown $r_s$ and 
$\Gamma$, the above estimates can only be improved with
future observations.

The unpolarized emission component dominates at $\sim 7$ mm,
where Lo et al. (1998) measure a source size of 
$\sim 5\times 10^{13}$ cm.  The self-absorption frequency
then requires $\gamma\sim 400$, 
$B \sim 0.1$ G, and $n_e \sim 4 \times 10^5$ cm$^{-3}$.  Though 
somewhat ad-hoc, this model reproduces the spectrum well.
The Faraday rotation parameter is rather small, so depolarization
requires field which is tangled on a scale $\sim 100$ times 
smaller than the size of the emission region.

\subsection{Accretion Component}

We have tried modeling the spectrum of the unpolarized component with a 
self-similar, self-absorbed accretion flow. 
We used the cyclo-synchrotron emission formulae from Mahadevan, Narayan, \& Yi (1996) and 
we performed the radiation transfer in full general relativity (Kurpiewski
\& Jaroszy\'nski 1997).  
We can place an upper limit on the accretion rate of an ADAF component (using 
the model of \"Ozel et al. 2000) since its unpolarized flux must not dilute 
the polarized component:  we find $\dot M_{ADAF} \la 3\times 10^{-6} M_\odot/$yr.
If the ADAF surrounds the polarized emission region, then it will depolarize,
so the Faraday depolarization places a stronger upper limit (\S 2).  We can 
place a similar limit on the CDAF model (using the structure from Quataert \& 
Gruzinov, 2000, with equipartition $B$ field and $p_{gas} = 2n_e k_B T_e$):  we 
find $\dot M_{CDAF} \la 1.5\times 10^{-9}M_\odot/$yr;  this accretion rate can account for 
the unpolarized component at $\nu \ga 10$GHz (see Figure 1) and is consistent 
with the Faraday rotation constraint.  The CDAF luminosity is $2\times 
10^{34}$erg/s and the self-absorption frequency is $\sim 30$GHz, so the polarized 
component would be visible through it.  Finally, we can place a limit on
a standard thin disk from the infrared upper limits:  we find $\dot M_{thin}
\la 2\times 10^{-11}M_\odot/$yr; this upper limit can be increased to
a maximum of $3\times 10^{-7} M_\odot/$yr if the inner edge of the disk is 
truncated at $r=6000 r_g$.

\section{Conclusions}

The main success of advection-dominated accretion models for Sgr A*
is in explaining the high-frequency radio spectrum and skirting
below the upper limits at infrared frequencies. 
However, the ADAF model is unpolarized at the same high frequencies, 
inconsistent with the recent detection of linear polarization.  We have 
constructed a simple toy model for the millimeter polarization which 
predicts a rise towards shorter
wavelengths:  polarization of $\sim 70\%$ might be seen with
SCUBA at 350 $\micron$ if this model is correct.  The lack of polarization
and spectral slope of 1/3 at wavelengths longer than 2 mm indicates
that a different physical component may be contributing.  The presence
of two physical components can be confirmed by looking for a change
in variability amplitude and time-scale or source size and morphology 
around 2 mm.

The high observed polarization implies a highly ordered magnetic field
lying near the sky plane.  This might be due to the poloidal field in 
a jet (Falcke, Mannheim, \& Biermann 1993), or
due to a toroidal field in a disk component seen edge-on.
The non-thermal electron distribution might be produced by shock
acceleration, reconnection, or electric field acceleration near 
the event horizon of a spinning black hole (Blandford \& Znajek
1977).  The Blandford-Znajek mechanism can generate a maximum 
luminosity of $L_{BZ} \sim 10^{37} (B/600 G)^2$ erg/s
(Thorne, Price, \& MacDonald 1986), so the entire polarized luminosity 
of Sgr A* might be powered by black hole spin.

The dynamics of the emission region will be controlled by the ratio of
the magnetic field energy density to the matter energy density, 
$B^2/(8\pi\rho c^2)$; however, this ratio scales as $r_s^{17} \Gamma^{9/2}$, 
while $\Gamma$ and $r_s$ are unknown.  Doppler boosting decreases the 
brightness temperature, which reduces Faraday rotation but makes the 
electrons trans-relativistic.  Future sub-mm VLBI observations should 
accurately measure the $r_s$ as a function of frequency, and proper 
motions may constrain $\Gamma$.  Also uncertain 
are the pair fraction and minimum electron energy $\gamma_{min}$.
The pair number density can be constrained by measuring the circular
polarization; without pairs, the circular polarization may be as high as 
a few percent at optically-thin wavelengths (Jones \& O'Dell 1977), while 
pure pair emission should have no circular polarization.  The pair 
annihilation line should be looked for at higher spatial resolution; 
however, it will be strongly broadened by relativistic motions of the pairs.
Once the source size is known, $\gamma_{min}$ and the pair fraction will be 
constrained by the Faraday rotation limit.

An ADAF model must have a low accretion rate, $\la10^{-8} M_\odot/$yr, to
be consistent with the lack of Faraday rotation of the polarized emission.
Such a low inferred accretion rate disagrees with estimates 
of the Bondi accretion rate inferred from stellar winds near the region 
of the black hole.  If accretion is episodic due to outer-disk 
instabilities, then the current state might be one of low accretion 
rate in the inner disk.  Alternatively, the accretion rate might be 
reduced by depositing energy from the accretion flow in the 
surrounding gas (either through outflow or convection), thus
increasing the sound speed and decreasing the capture rate of gas
by the black hole.  The accretion flow must deposit energy 
$\dot M_A GM/r_A \sim 6\times 10^{35}$ erg/s, where $\dot M_A$ is the stellar 
mass loss rate which crosses the Bondi radius $r_A$ (Quataert et al. 1999).  
This can be supplied by accretion
which releases energy $\sim 5\times 10^{35}(\eta/0.01) \dot m_{-9} $ erg/s,
where $\eta$ is the efficiency with which accretion deposits energy at large
radius.  As remarked above, a 
convection-dominated accretion flow with $\dot M \sim 10^{-9}
M_\odot/$yr can explain part of the unpolarized component without diluting the 
polarized emission;  the associated convection can carry the required 
energy outward to suppress the Bondi accretion rate.

Since the self-absorption frequency occurs at $\sim 500\micron$, it will 
be possible to image shadow of a black hole from the ground using VLBI, 
providing a direct confirmation of the existence of an event horizon (Falcke, 
Melia, \& Agol 2000).  Future sub-mm polarimetric VLBI observations might 
show rotation of the polarization angle near the black hole, a general 
relativistic effect which becomes stronger for a spinning black hole 
(Connors, Stark, \& Piran 1980). 

Eliot Quataert \& Andrei Gruzinov have shown me work which reaches
similar conclusions about ADAFs and the need for a very low accretion
rate onto Sgr A*.

\acknowledgments

I acknowledge Ski Antonucci, Julian Krolik, Colin Norman, and
Eliot Quataert for ideas and corrections which greatly improved 
this letter.  This work was supported by NSF grant AST 96-16922.


\begin{references}
Agol, E. \& Blaes, O.M., 1996, MNRAS, 292, 965

Aitken, D. K., et al., 2000, ApJ, in-press, astro-ph/0003379, A00

Bardeen, J.~M. 1973, in { Black Holes}, ed. C. DeWitt \& B. S. DeWitt
(New York: Gordon \& Breach), 215

Begelman, M. C. \& Blandford, R. D., 1999, MNRAS, 303, L1

Bower, G.~C., Backer, D.~C., Zhao, J.~H., Goss, M., \& Falcke, H.  1999a, ApJ, 521, 582

Bower, G.~C., Wright, M. C.~H., Backer, D.~C., \& Falcke, H. 1999b, ApJ, 527, 851

Beckert, T. \& Duschl, W. J., 1997, A\& A, 328, 95

Blandford, R. D. \& Znajek, R. L, 1977, MNRAS, 179, 433

Coker, R. \& Melia, F., 1999, ApJ, 511, 750

Connors, P. A., Stark, R. F., \& Piran, T., 1980, ApJ, 235, 224

Falcke, H., Mannheim, K., \& Biermann, P. L., 1993, A\&A, 278, L1

Falcke, H., Melia, F., \& Agol, E., 2000, ApJ, 528, L13

Genzel, R., Eckart, A., Ott, T., \& Eisenhauer, F., 1997, MNRAS, 291, 219

Ghez, A.~M., Klein, B.~L., Morris, M., \& Becklin, E.~E. 1998, ApJ, 509, 678

Ginzburg, V. L. \& Syrovatskii, S. I., 1965, ARA\&A, 3, 297, GS

Ginzburg, V. L. \& Syrovatskii, S. I., 1969, ARA\&A, 7, 375, GS

Igumenshchev, I. V. \& Abramowicz, M. A., 1999, MNRAS, 303, 309

Igumenshchev, I. V. \& Abramowicz, M. A., 2000, astro-ph/003397

Igumenshchev, I. V.,  Abramowicz, M. A., \& Narayan, R., 2000, astro-ph/004006

Krichbaum, T.~P. et al.  1998, A\&A, 335, L106

Lo, K.~Y., Shen, Z.~Q., Zhao, J.~H., \& Ho, P. T.~P. 1998, ApJ, 508, L61

Kurpiewski, A. \& Jaroszy\'nski, M., 1999, A\&A, 346, 713

Jones, T. W. \& Hardee, P. E., 1979, ApJ, 228, 268

Jones, T. W. \& O'Dell, S. L., 1977, ApJ, 214, 522

Mahadevan, R., 1999, MNRAS, 304, 501

Mahadevan, R., Narayan, R., \& Yi, I., 1996, ApJ, 465, 327

Melia, F., 1992, ApJ, 387, L25

Narayan, R. \& Yi, I., 1994, ApJ, 428, L13

Narayan, R., Yi, I., \& Mahadevan, R., 1995, Nature, 374, 623

Narayan, R., Igumenshchev, I. V. \& Abramowicz, M. A., 2000, astro-ph/9912449

\"Ozel, F., Psaltis, D., \& Narayan, R., 2000, ApJ, in press, astro-ph/0004195

Quataert, E., Narayan, R., \& Reid, M. J., 1999, ApJ, 517, L101

Quataert, E. \& Narayan, R., 1999, ApJ, 520, 298

Quataert, E. \& Gruzinov, A., 2000, ApJ, in press, astro-ph/9912440

Stone J. M., Pringle J. E., \& Begelman M. C., 1999 MNRAS, 310, 100

Thorne, K. S., Price, R. H., \& Macdonald, D., 1986, Black
Holes: The Membrane Paradigm. Yale Univ. Press, New Haven, 
CN

\end{references}
\end{document}